\documentclass[]{aastex631}
\usepackage{multirow}
\usepackage{xspace}

\usepackage{amsmath}

\newcommand{\Msun}{\,{\rm M_\odot}}
\newcommand{\fedd}{\,{f_{\rm Edd}}}
\newcommand{\Mblack}{M_\bullet}
\newcommand{\Mstar}{M_\star}

\newcommand{\mmstar}{$\Mblack-\Mstar$\xspace}

\accepted{\today}

\submitjournal{Research Notes of the AAS}

\begin{document}

\title{The Host Galaxy of a Dormant, Overmassive Black Hole at $z=6.7$ May Be Restarting Star Formation}

\correspondingauthor{Fabio Pacucci}
\email{fabio.pacucci@cfa.harvard.edu}

\author[0000-0001-9879-7780]{Fabio Pacucci}
\affiliation{Center for Astrophysics $\vert$ Harvard \& Smithsonian, Cambridge, MA 02138, USA}
\affiliation{Black Hole Initiative, Harvard University, Cambridge, MA 02138, USA}

\author[0000-0003-4330-287X]{Abraham Loeb}
\affiliation{Center for Astrophysics $\vert$ Harvard \& Smithsonian, Cambridge, MA 02138, USA}
\affiliation{Black Hole Initiative, Harvard University, Cambridge, MA 02138, USA}

\author[0009-0003-7423-8660]{Ignas Juod\v{z}balis}
\affiliation{Kavli Institute for Cosmology, University of Cambridge, Cambridge, CB3 OHA, UK}
\affiliation{Cavendish Laboratory, University of Cambridge, Cambridge, CB3 OHE, UK}



\begin{abstract}
JWST is discovering a large population of $z>4$ supermassive black holes (SMBHs) that are overmassive with respect to the stellar content of their hosts. A previous study developed a physical model to interpret this overmassive population as the result of quasar feedback acting on a compact host galaxy. In this Note, we apply this model to JADES GN 1146115, a dormant supermassive black hole at $z=6.7$ whose mass is $\sim40\%$ of the host's mass in stars and accreting at $\sim2\%$ of the Eddington limit. The host has been forming stars at the low rate of $\sim 1 \rm \Msun\,yr^{-1}$ for the past $\sim 100$ Myr. Our model suggests that this galactic system is on the verge of a resurgence of global star formation activity. This transition comes after a period of domination by the effect of its overmassive black hole, whose duration is comparable to typical quasar lifetimes.  
\end{abstract}


\section{Introduction} 
\label{sec:intro}
The James Webb Space Telescope (JWST) has detected a large population of supermassive black holes (SMBHs) at $z>4$ with typical masses in the range $10^6-10^8 \Msun$ (see, e.g., \citealt{Harikane_2023, Maiolino_2023_new}).
Remarkably, the ratio between black hole mass $\Mblack$ and stellar mass $\Mstar$ is significantly higher than the local value of $\sim 10^{-3}$ \citep{Reines_Volonteri_2015}.
With a detailed statistical analysis, \cite{Pacucci_2023_JWST} concluded that the population of galaxies currently probed by JWST host SMBHs that are $10-100$ times overmassive with respect to their stellar content, in comparison with the local population of galaxies, and derived a high-$z$ \mmstar relation. Notably, a virtually indistinguishable relation was found, independently, by \cite{Inayoshi_2024_spin}.

Significant uncertainties characterize the determination of $\Mblack$ and $\Mstar$, which rely on methods calibrated at low redshift. \cite{Pacucci_2023_JWST} pointed out that the tension with the local \mmstar relation will vanish if there is a systematic error of a factor $\sim 60-70$ in the measurements of the \textit{ratio} between black hole mass and stellar mass, in a way such that the black hole (stellar) masses are overestimated (underestimated). As these galaxies are physically small (i.e., they have effective radii of $r_e \sim 150$ pc, see \citealt{Baggen_2023}), it is unlikely that their stellar masses are underestimated by significant factors. 

\cite{Pacucci_Loeb_2024} developed a framework to explain this population of overmassive SMBHs at $z>4$, based on two assumptions. First, these galaxies are small, so the central SMBH has an outsized thermal effect on the entire host. Second, for $z>4$, the SMBH's growth times are comparable to the age of the Universe; hence, they are constantly injecting energy into the system and heating the gas. 
Once the central SMBH's duty cycle drops significantly below unity, the gas can finally cool down, and effective global star formation can resume. In this framework, \cite{Pacucci_Loeb_2024} derived a condition for the average star formation efficiency to be quenched.

In this Research Note, we directly apply this framework to JADES GN 1146115, an extremely overmassive black hole at $z=6.7$ \citep{Juodzbalis_2024}.

\section{Results}
\label{sec:results}
JADES GN 1146115 is a compact galaxy ($r_e \approx 140$ pc) of stellar mass $\log{(\Mstar/\Msun)} = 8.92^{+0.30}_{-0.31}$, hosting a black hole with mass $\log{(\Mblack/\Msun)} = 8.61^{+0.27}_{-0.24}$, thus leading to a $\Mblack/\Mstar$ ratio of $\sim 40\%$. Its Eddington ratio is $\fedd =0.024^{+0.011}_{-0.008}$, with a systematic scatter of $0.5$ dex, making it an underluminous, ``dormant'' SMBH.

\begin{figure*}%
    \centering
\includegraphics[angle=0,width=0.85\textwidth]{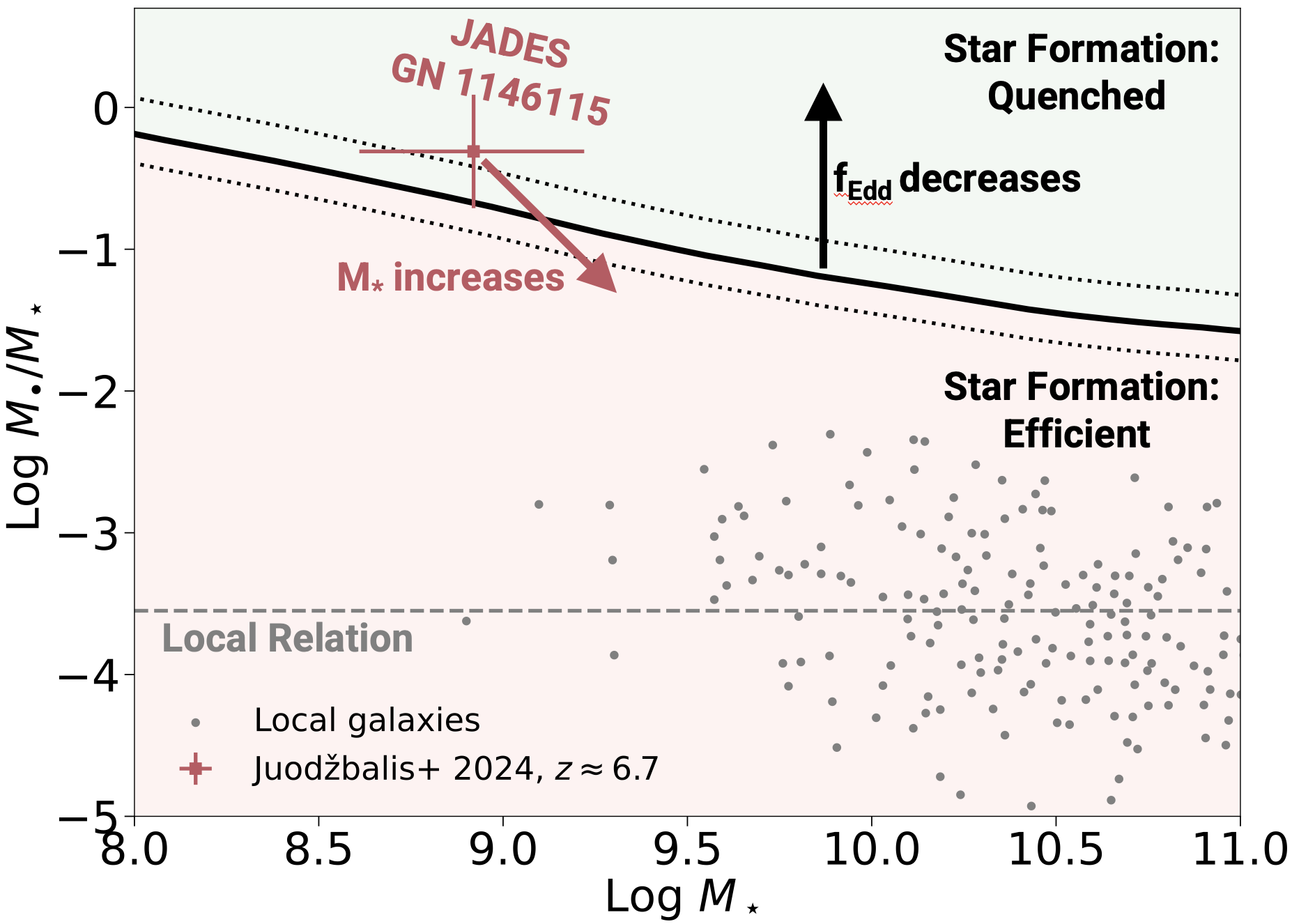} \hfill
    \caption{The condition for SMBH-driven star formation quenching developed in \cite{Pacucci_Loeb_2024} is applied to the galaxy JADES GN 1146115 \citep{Juodzbalis_2024}, shown in the [$\Mstar$, $\Mblack/\Mstar$] plane with its $1\sigma$ uncertainties. The dotted lines indicate how the range of $\fedd$ estimated for JADES GN 1146115 affects the threshold on $\Mblack/\Mstar$. Galaxies hosting overmassive black holes in the green area experience reduced star formation as quasar feedback increases the gas temperature above the virial one. Local galaxies on the \mmstar relation (shown as a dashed line) are indicated with gray symbols \citep{Reines_Volonteri_2015}. The red arrow indicates the effect of an increasing host's stellar mass in changing the location of the galaxy in the plane; the black arrow indicates the effect of a decreasing $\fedd$ in raising the threshold value of $\Mblack/\Mstar$. Both changes would lead JADES GN 1146115 to a region where star formation is more efficient.}
    \label{fig:condition}%
\end{figure*}

For a detailed description of the theoretical model, the interested reader is referred to \cite{Pacucci_Loeb_2024}.
In summary, star formation in a compact, high-$z$ galaxy is quenched if the SMBH injects enough thermal energy into the host to raise its gas temperature above the virial one.
Expressing this condition in terms of the ratio $\Mblack/M_\star$:
\begin{equation}
    \frac{\Mblack}{\Mstar} > 8 \times 10^{18} \frac{n \Lambda}{\fedd} \left( \frac{\Omega_b}{\Omega_m}\frac{M_h}{\Mstar} - 1 \right) \, .
\end{equation}
Here, $n$ is the average gas number density, $\Lambda$ is the cooling function, $\Omega_b/\Omega_m$ is the baryon fraction, and $M_h$ is the halo mass.
Compared to \cite{Pacucci_Loeb_2024}, here we only adapt our assumption on $\fedd$. Instead of assuming $\fedd = 1$ (as JWST's overmassive systems are found to have rates $0.1 < \fedd < 5$, see, e.g., \citealt{Harikane_2023, Maiolino_2023_new}), we use the actual $\fedd$ measured by \cite{Juodzbalis_2024}, with its $1\sigma$ uncertainty: $\fedd =0.024^{+0.011}_{-0.008}$.

The location of JADES GN 1146115 in the [$\Mstar$, $\Mblack/\Mstar$] plane is shown in Figure \ref{fig:condition}. In summary, JADES GN 1146115 is at the threshold between regions of efficient and inefficient star formation.

\section{Interpretation}
\label{sec:interpretation}
JADES GN 1146115 is a galaxy characterized by an instantaneous star formation rate (SFR) of  $1.38^{+0.92}_{-0.45} \, \rm \Msun \, yr^{-1}$, lower by a factor of $\sim 3$ compared to similar galaxies. Furthermore, this rate has been nearly constant in the last $\sim 100$ Myr \citep{Juodzbalis_2024}. 
This time scale is intriguingly similar to typical quasar lifetimes, i.e., $10^7-10^8$ yr, which is the duration of rapid black hole growth \citep{Hopkins_2005}.

Our physical interpretation of the galactic system JADES GN 1146115, based on the \cite{Pacucci_Loeb_2024} model, is the following. Its central SMBH  grew at high rates for a typical quasar lifetime, overgrowing the stellar content of its host and pushing the galaxy well inside the region in the [$\Mstar$, $\Mblack/\Mstar$] plane where star formation is globally inefficient. Possibly due to a decrease in the stable gas supply, the central SMBH is now in a dormant state, accreting at low levels. The reduced effect of quasar feedback decreases the temperature floor, thus effectively placing the system at the transition boundary in the [$\Mstar$, $\Mblack/\Mstar$] plane. As shown in Fig. \ref{fig:condition}, two events may eventually push the system deep into the region where global star formation is efficient. First, an additional decrease in the Eddington ratio. Second, a stable (albeit slow) increase in the stellar content of the galaxy; at the current SFR, this galaxy needs $\sim 1$ Gyr to double its mass \citep{Juodzbalis_2024}. Any combination of these events would allow the system to restart efficient star formation, possibly closer to the star-forming main sequence (e.g., \citealt{Popesso_2023}).

In summary, the application of our model to JADES GN 1146115 suggests that we may be witnessing the slow resurgence of efficient star formation in a galaxy that has been dominated by its extremely overmassive black hole in the past $\sim 100$ Myr.

\bibliography{ms}
\bibliographystyle{aasjournal}



\end{document}